\documentclass[prb,showpacs,twocolumn,preprintnumbers,amsmath,amssymb]{revtex4}

\usepackage{graphicx}
\usepackage{epsfig}
\usepackage{dcolumn}% Align table columns on decimal point
\usepackage{bm}% bold math

\begin{document}

\title{Practical design and simulation of silicon-based quantum dot qubits}

\author{Mark Friesen$^{1,2}$}
\email{friesen@cae.wisc.edu}
\author{Paul Rugheimer$^2$}
\author{Donald E.~Savage$^2$}
\author{Max G.~Lagally$^{1,2}$}
\author{Daniel W.~van der Weide$^3$}
\author{Robert Joynt$^1$}
\author{Mark A.~Eriksson$^1$}
\email{maeriksson@facstaff.wisc.edu}
\affiliation{$^1$Department of Physics, University of Wisconsin, Madison, 
Wisconsin 53706 \\
$^2$Department of Materials Science and Engineering, University of Wisconsin, Madison, 
Wisconsin 53706 \\
$^3$Department of Electrical and Computer Engineering, University of Wisconsin, Madison, 
Wisconsin 53706}

\begin{abstract}
Spins based in silicon provide one of the most promising architectures for quantum computing.  A scalable design for silicon-germanium quantum dot qubits is presented. The design incorporates vertical and lateral tunneling.  Simulations of a four-qubit array suggest that the design will enable single electron occupation of each dot of a many-dot array.  Performing two-qubit operations has a negligible effect on other qubits in the array.  Simulation results are used to translate error correction requirements into specifications for gate-voltage control electronics.  This translation is a necessary link between error correction theory and device physics.
\end{abstract}


\maketitle

Quantum computation would enable huge speedups of certain very hard problems, such as factorization.\cite{shor94}  However, quantum computing is essentially an analog method.  As such, the problem of errors creates a serious challenge.  Advances in error correction algorithms have produced well-justified optimism that this challenge will be overcome.\cite{preskill98}  However, existing error correction algorithms require low error rates.   Thus, hardware design will be critical to the creation of a working quantum computer.

The purpose of this paper is to address hardware design challenges in a specific materials system: silicon quantum dots.  There are two reasons to analyze this system in detail.   First, spins in silicon have long coherence times.\cite{note1}  Second, classical silicon electronics has demonstrated fast operation and a proven record of scalable integration. Indeed, several spin-based qubit designs have emerged that are compatible with silicon.\cite{loss98,kane98,vrijen00,mozyrsky01,levy01}  The full benefit of existing silicon technology may be used to greatest advantage in spin qubits based in quantum dots.  Previous calculations of the exchange coupling in coupled quantum dots with idealized potentials have demonstrated the promise of such structures for quantum computation.\cite{burkard99,hu00}  However, there are important questions that can only be addressed in the context of an explicit physical design and realistic simulations.

In this paper, we present an explicit design for quantum dot qubits in silicon-germanium heterostructures.  To determine whether it will be possible to build and operate such a device, we perform realistic simulations of four coupled qubits.  The simulations are self-consistent:  they include the full three-dimensional electrostatics, and the Hamiltonian is solved via exact diagonalization.\cite{methods}  These simulations allow us to answer several questions.  First, we find that it is possible to couple neighboring quantum dot qubits without any significant perturbation of secondary qubits.  Second, the coupling can be strong, enabling GHz operation rates.  Most importantly, these simulations allow us to translate gate voltage uncertainties---which are inevitable---into error rates in quantum gates.  This translation is the necessary link between device physics and quantum error correction theory.

In this paper, we do not propose a new scheme for quantum computation.  Rather, we perform simulations of a new design suitable for implementing the scheme of Loss and DiVincenzo.\cite{loss98}  The quantum computer we have in mind is the following: the physical qubits are individual electron spins in quantum dots.  Two-qubit operations are performed on these physical qubits by controlling the exchange coupling $J$ as a function of time.  Logical qubits can be coded into a subspace of the physical qubits, so that the exchange coupling alone enables universal quantum computation.\cite{bacon00,divincenzo00}  Initialization of the coded qubits is performed according to the scheme of DiVincenzo {\it et al}.\cite{divincenzo00}  Readout is performed via spin-charge transduction, as in the tunneling scheme of Kane.\cite{kane98}
\begin{figure}
\centerline{\epsfxsize=2.4in \epsfbox{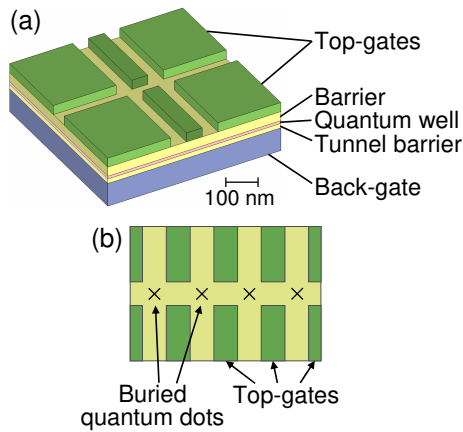}}
\caption{
The two quantum dot devices simulated in this paper.  (a)  A double-dot structure, as studied in Figs.~2 and 3. From bottom to top, the heterostructure cross section is composed of a thick, $n$-doped, strain-relaxed Si$_{1-x}$Ge$_x$ back-gate, a 10~nm undoped Si$_{1-x}$Ge$_x$  tunnel barrier, a 6~nm undoped Si quantum well, a 20~nm undoped Si$_{1-x}$Ge$_x$ barrier, and lithographically-patterned metallic top gates.  All fabrication steps are based on standard technology.  Not pictured is a thin Si capping layer.  (b) A four-dot structure, as studied in Fig.~4. [Top-view only;  heterostructure identical to (a).]  Periodic boundary conditions are assumed.  The dots reside in the quantum well layer, at positions marked by $\times$'s.}
\end{figure}

The quantum computer just described is well defined, but it is abstract.  Our specific design is shown in Fig.~1(a).  It incorporates aspects of two existing types quantum dots; lateral tunneling dots and vertical tunneling dots.\cite{kouwenhoven97}  The quantum dot of Fig.~1(a) is defined by a quantum well that confines electrons vertically, and by split top gates that confine electrons laterally, by electrostatic repulsion.  These features are typical of lateral quantum dots.  The device of Fig.~1(a) differs from a typical lateral quantum dot because it contains a tunnel-coupled back gate, usually found only in vertical quantum dots.  As the simulations below show, the back gate allows tuning of the electron number in each quantum dot, even when those quantum dots are part of a large array.  The back gate also screens the Coulomb interaction.  All semiconductor layers in this design are composed of strain-relaxed Si$_{1-x}$Ge$_x$ except the quantum well, which is formed of strained silicon.  Relaxation in SiGe can be achieved by step-graded compositional growth on a substrate silicon wafer.\cite{mooney95}  In the simulations presented here, we use the composition $x=0.077$, consistent with a quantum well band offset of 
$\Delta E_c \simeq 84$~meV with respect to the barriers.
Since the transverse effective mass and dielectric constant change little with $x$ when $x$ is small, we use constant 
values $m_t=0.19m_e$ and $\varepsilon=11.9\varepsilon_0$ throughout the heterostructure.

The zeroth order requirement for an individual electron qubit is that it should contain an individual electron.  Fig.~2 shows the stability energy (the energy cost to change the electron number) calculated for the pair of coupled quantum dots shown in Fig.~1(a).  The results are plotted as a function of the gate voltages $V_{\rm in}$ and $V_{\rm out}$, as explained in Fig.~3 (inset).  The stability energy is greater than 1 meV over a wide range of gate voltages.  We have also performed stability calculations for the four-qubit device of Fig.~1(b), and we find stability greater than 1~meV throughout the operating range discussed below.

Qubits are useful only if operations can be performed on them.  For the structure we describe here, the operations are performed by controlling the exchange coupling $J(t)$ between neighbor qubit pairs, with the interaction Hamiltonian 
$H_s(t)=J(t){\bf S}_1 \cdot {\bf S}_2$.  The exchange coupling is large only when the electron wavefunctions overlap.  It can be made exponentially small by forcing the electrons to separate.  These manipulations are performed via the top gate voltages, and these gate voltages translate directly into the time evolution of the qubits.
The mapping $J(V_1,V_2,\ldots )$ between the exchange coupling and the top-gate voltages is an operational characterization of the quantum computer.  As we show below, knowledge of this mapping allows us to determine error rates for our design.  Here, we calculate the exchange coupling as the energy difference between the ground and first excited states:\cite{burkard99,hu00} 
$J=E_{\rm trip}-E_{\rm sing}$, where ``singlet" and ``triplet" refer to the spin symmetry of the two-electron wavefunction.  Significant numerical accuracy is required in the calculations, because of the large difference in energy scales:
$J/E_{\rm trip}<5\times 10^{-4}$.
\begin{figure}[b]
\centerline{\epsfxsize=2.3in \epsfbox{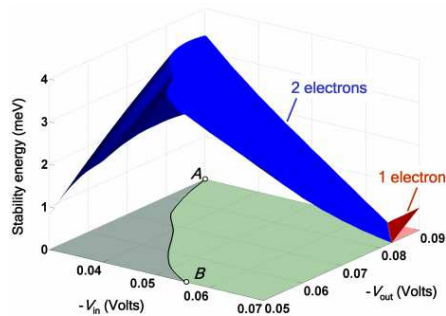}}
\caption{
The stability energy (energy to change the electron filling number in a two-qubit device) {\it vs.}\ gate voltages, computed for the two-qubit device, Fig.~1(a).  The two-electron stability range is shown in the center, while the one-electron stability range is shown on the right.  Three-electron stability does not occur in the voltage range shown here.  Optimal two-electron stability is obtained along curve $AB$.}
\end{figure}

Figure 3 shows a map of the exchange coupling $J$ as a function of gate voltages 
$V_{\rm in}$ and $V_{\rm out}$, computed numerically for the double-dot device of Fig~1(a).  The back-gate is grounded.  
The envelope function approximation used here is reasonable for quantum dots of size $\sim 50$~nm.\cite{note6}
The overall trends in Fig.~3 are consistent with previous studies, which use more idealized confinement potentials.\cite{burkard99,hu00}  However, because the magnetic field is zero in this work, the exchange coupling does not cross zero, in contrast with results for high magnetic fields.  Nonetheless, $J$ can be made arbitrarily small by raising the electrostatic barrier between the qubits.  Raising such a barrier creates an asymptotic approach to zero that is extremely robust.
\begin{figure}
\centerline{\epsfxsize=2.3in \epsfbox{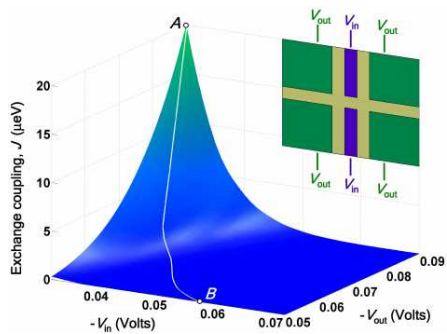}}
\caption{
Exchange coupling $J$ {\it vs.}\ gate voltages, computed for the two-qubit device, Fig.~1(a).  Top gate voltages $V_{\rm out}$ and $V_{\rm in}$ are described in the inset, while the back gate is held to ground.  Curve $AB$ marks the line of maximum stability for gate operation. (See Fig.~2.)}
\end{figure}

The data in Figures 2 and 3 are for two qubits.  It is conceivable that adding additional qubits would cause at least one qubit to become unstable or suffer an undesired coupling with a neighboring qubit during the manipulations described by Fig.~3.  Fortunately, this is not the case.  Fig.~4 shows the electron density in a four-qubit device for two extreme cases, in which the exchange coupling is either (a) very small or (b) very large.  Between Fig.~4(a) and (b), the inner pair of electrons each move by 21.5~nm, whereas the outer electrons move by only 0.5 nm.  This motion corresponds to a change in $J$ for the inner pair from approximately $10^{-19}$~eV to 0.4~$\mu$eV.  We can estimate the $J$ coupling between the fourth and ``fifth" electrons in Fig.~4 (using periodic boundary conditions).  We obtain approximately $10^{-19}$~eV for Fig.~4(a), and this number decreases by only a factor of 0.8 in Fig.~4(b).  Thus, any pair of qubits can be manipulated independently of any other pair.  This independence is due, in part, to screening effects arising from the various gates.
\begin{figure}[b]
\centerline{\epsfxsize=2.3in \epsfbox{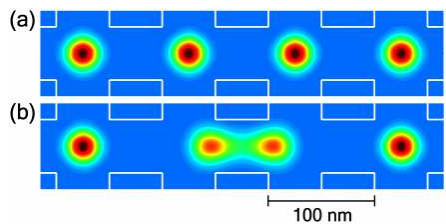}}
\caption{
Simulations of the four-qubit device, Fig.~1(b) (gates outlined in white).  (a) Coupling ``off," with $J\simeq 10^{-19}$~eV (top-gate voltages all set to -0.15~V).  (b) Coupling ``on," with $J=0.4$~$\mu$eV (center gate voltage set to -0.075~V).}
\end{figure}

The results of Fig.~3 allow us to consider errors in quantum gates.  It is important to remember that errors arise in quantum computing not just from decoherence but also from the inevitable misapplication of quantum gates.  Such misapplications will arise, for example, from uncertainties in the applied gate voltages.  Fault-tolerant techniques have been developed for correcting errors, but these are only effective below an error threshold of one accumulated error in $10^4$ operations.\cite{note2}  Thus, it is critical to know the error rate expected during the application of quantum gates.
	
Here we calculate the error rate in $J$ as a function of the uncertainty or noise in the voltage pulses used to manipulate the quantum dots.  Accurate gate control involves two steps: (i) initial characterization of the exchange coupling between pairs of qubits, and (ii) precise implementation of the gate operations.  For this discussion, we assume perfect characterization, and we focus on step (ii).  As a prototype for gate operations we consider $\sqrt{\rm SWAP}$, as implemented with a voltage pulse $V_s(t)$.  In principle, the particular shape of $V_s(t)$ is arbitrary, although it must satisfy the following relation:\cite{loss98}
\begin{equation}
\int_{\tau_s} J[V_s(t)]\, dt = \pi \hbar /2 .
\end{equation}
Here, $\tau_s$ is the switching time, and the function $J(V)$ was computed in Fig.~3.  To be specific, we consider low and flat voltage pulses, such that errors in the pulse width are diluted to acceptable levels.\cite{note3}

What are the error levels that can be tolerated in the applied gate voltages?  For a flattop pulse of height 
$J=\pi \hbar /2\tau_s$, fault tolerant computation requires that the pulse height uncertainty $\delta V$ should satisfy
\begin{equation}
\delta V < 10^{-4}J \left| \frac{\partial J}{\partial V} \right|^{-1} .
\label{eq:fault}
\end{equation}
It is important to note that the magnitude of $\delta V$ depends on classical control electronics, while $J$ and $\partial J/\partial V$ are implementation-specific.  To evaluate gating errors, it is therefore necessary to work with a realistic device design.  By fitting the exponential dependence of $J(V)$ in Fig.~3, Eq.~(\ref{eq:fault}) yields the requirement 
$\delta V/V<5$-$8\times 10^{-6}$ for the double qubit device of Fig.~1(a).  We can compare this figure to published specifications for state-of-the-art control electronics.  For sub-kHz pulses (approaching DC), extremely high voltage accuracy can be achieved, and the requirement can be met.  For sub-MHz pulse generators, the desired accuracy levels fall nearly within the specifications of off-the-shelf electronics.\cite{note4}  For GHz operation, over three orders of magnitude improvement in pulse height uncertainties will be required to meet the requirements of fault tolerant computation.\cite{note5}  We point out that decoherence constraints may indeed require spin-based silicon qubits to operate in the GHz regime.

In conclusion, we have described and simulated a realizable design for a SiGe quantum dot quantum computer.  A prominent feature of this device is the back-gate, which enables tuning of the number of electrons in each quantum dot.  We have directly addressed the issue of scalability through simulations of a four-qubit device.  Qubit interactions are found to be very robust, particularly as a consequence of Coulomb screening provided by the back-gate.  Our calculations show that a key challenge for solid-state spin-based quantum computation is to develop devices in which the exchange coupling is relatively insensitive to gate voltage uncertainty.  At a simple level, the quantum dot structures should be optimized to increase $J/|\partial J/\partial V|$, which sets the scale for gate voltage accuracy requirements.  The ultimate goal should be to ``digitize" the gating function $J(V)$, such that $\partial J/\partial V$ goes to zero at appropriate working points.\cite{friesen02}

%\vspace{.3in}
We have benefited from helpful discussions with C.~L.~Brace, S.~Coppersmith, X.~Hu, D.~A.~Lidar, and C.~Tahan.  R.~Nelson and E.~Blevis provided invaluable technical support with the PDE modeling software, FlexPDE$^\copyright$.  Our work was supported by the U.S.\ Army Research Office through the ARDA program, and the National Science Foundation through the MRSEC and QuBIC programs.


\begin{thebibliography}{99}

\bibitem{shor94}
P. W. Shor, in {\it Proc. 35th Annual Symposium on Foundations of Computer Science}, edited by S. Goldwasser (IEEE Computer Society Press, Los Alamitos, CA, 1994), pp. 
124-134.

\bibitem{preskill98}
J. Preskill, Proc. R. Soc. London A {\bf 454}, 385 (1998).

\bibitem{note1}
For donor-bound electrons in bulk Si, for example, the spin-lattice relaxation time $T_1$ can be greater than 3000~s at low temperatures and fields.\cite{feher59}  When uniaxial strain is applied, $T_1$  grows by many orders of magnitude.\cite{tahan02}  The transverse spin relaxation time $T_2^*\simeq 0.5$~ms is typically shorter, and can be enhanced through the use of isotopically enriched $^{28}$Si to minimize the effect of nuclear spins.\cite{gordon58}  It is generally expected that decoherence times in confined structures like quantum dots will exceed the bulk values.

\bibitem{loss98}
D. Loss and D. P. DiVincenzo, \pra {57}, 120 (1998).

\bibitem{kane98}
B. E. Kane, Nature {\bf 393}, 133 (1998).

\bibitem{vrijen00}
R. Vrijen, E. Yablonovitch, K. Wang, H. W. Jiang, A. Balandin, V. Roychowdhury, T. Mor, and D. DiVincenzo, \pra {\bf 62}, 012306 (2000).

\bibitem{mozyrsky01}
D. Mozyrsky, V. Privman, and M. L. Glasser, \prl {\bf 86}, 5112 (2001).

\bibitem{levy01}
J. Levy, \pra {\bf 64}, 052306 (2001).

\bibitem{burkard99}
G. Burkard, D. Loss, and D. P. DiVincenzo, \prb {\bf 59}, 2070 (1999).

\bibitem{hu00}
X. Hu and S. Das Sarma, \pra {\bf 61}, 062301 (2000).

\bibitem{methods}
The electrostatic and Hartree-Fock calculations were performed using finite-element software, FlexPDE$^\copyright$.  Image potentials arising from the nontrivial gate structure were calculated self-consistently using a numerical Green's function technique.  (Test charges were first introduced into the heterostructure.  The direct contributions from the test charges were then subtracted from the solutions to obtain exact image potentials.)  The quantum mechanical problem was approximated as a single envelope function.\cite{note6}  A basis set of 18 single-electron Hartree-Fock wavefunctions was obtained in real space on an adaptive finite-element mesh.  A basis of about 50 two-electron wavefunctions was then constructed in the configuration interaction approach.  The Hamiltonian matrix was computed and diagonalized, giving an essentially exact result for the envelope function.  For the four-qubit simulation (Fig.~1b), the wavefunctions of the outer two electrons were found to have an insignificant overlap with the two center electrons.  Their couplings were therefore treated as purely Coulombic.  It was not possible to calculate directly the tiny values of $J$ corresponding to high potential barriers between the quantum dots.  In this regime, we made use of the nearly exponential dependence of $J$ on the gate voltage to obtain extrapolations.

\bibitem{bacon00}
D. Bacon, J. Kempe, D. A. Lidar, and K. B. Whaley, \prl {\bf 85}, 1758 (2000).

\bibitem{divincenzo00}
D. P. DiVincenzo, D. Bacon, J. Kempe, G. Burkard, and K. B. Whaley, Nature {\bf 408}, 
339 (2000).

\bibitem{kouwenhoven97}
L. P. Kouwenhoven et al., in {\it Mesoscopic Electron Transport}, edited by 
L. L. Sohn, L. P. Kouwenhoven and G. Sch\"{o}n, (Kluwer Academic Publishers, Dordrecht, 1997) vol. 345, pp. 105-214.

\bibitem{mooney95}
P. M. Mooney et al., Appl. Phys. Lett. {\bf 67}, 2373 (1995).

\bibitem{note6}
Due to strain effects, the six-fold degeneracy of the silicon conduction band in the quantum well is lifted, so that only the $[001]$ bands are populated.
The envelope function approximation can be used to describe lateral variations of the wave function
when the confinement potential varies slowly with respect to atomic distances, for quantum dots of
size $\sim 50$~nm, the approximation should be reasonable.  
In the vertical direction, however,  
the band edge has sharp discontinuities at the quantum well interfaces.
For this case, an extension of Ref.~[\onlinecite{sham79}] to quantum wells shows that the envelope function approximation remains accurate, although the fine structure of the wavefunction (i.e., the Bloch function) should be modified to accomodate intervalley scattering.
Symmetry allows two possible valley-coupled wavefunctions:  
$\Psi_\pm =F_z(\phi_{+z}\pm \phi_{-z})$, 
where $\phi_{+z}$ and $\phi_{-z}$ are the appropriately modified Bloch functions at the $[001]$ and $[00\bar{1}]$ band minima.
The envelope functions in these two directions are equal, $F_{\pm z}\equiv F_z$.
Perturbation theory gives the energy splitting for the two eigenstates $\Psi_\pm$, although a complete treatment of the interface physics is still lacking.
Using the estimate given in Ref.~[\onlinecite{sham79}] ($\alpha\simeq0.5$~\AA) we predict a valley splitting of $\Delta E>0.06$~meV (0.7~K) for our heterostructure--more than 10 times our dilution fridge temperature.
By optimizing the heterostructure to increase the electric field in the quantum well, it is possible to increase $\Delta E$ significantly.
At low temperatures then, all electrons should be in the valley-split ground state, with no interference effects of the type discussed in Ref.~[\onlinecite{koiller02}].

\bibitem{note2}
The $10^{4}$ estimate assumes two-qubit operations between any pair of qubits.  In a linear qubit array with only nearest neighbor couplings, a more restrictive threshold may be appropriate.

\bibitem{note3}
Using the estimate $T_2>T_2^* \simeq 0.5$~ms for bulk silicon,\cite{gordon58} the fault-tolerant error threshold suggests a maximum pulse length of 0.1~$\mu$s.  For a flat-top pulse of width 0.1~$\mu$s, we estimate a minimum pulse edge of 10~ps  to satisfy adiabatic gating requirements.\cite{schliemann01,hu02}  Such a pulse must be produced with less than $10^{-4}$ relative error in its duration.  Such accuracy is currently beyond the limits of commercial pulse generation technology.

\bibitem{note4}
Estimates are based on specifications for pulse amplitude jitter in PB-4 and PB-5 sub-MHz pulse generators from Berkeley Nucleonics Corporation (http://www.berkeleynucleonics.com).

\bibitem{note5}
Specifications from the Agilent Technologies 8133 and 81100 families of GHz pulse generators are listed as 
$\delta V/V<0.01$ (http://www.agilent.com).

\bibitem{friesen02}
M. Friesen, R. Joynt, and M. A. Eriksson, \apl \textbf{81}, 5619 (2002).

\bibitem{feher59}
G. Feher and E. A. Gere, Phys. Rev. {\bf 114}, 1245 (1959).

\bibitem{tahan02}
C. Tahan, M. Friesen, and R. Joynt, \prb {\bf 66} 035314 (2002).

\bibitem{gordon58}
J. P. Gordon and K. D. Bowers, \prl {\bf 1}, 368 (1958).

\bibitem{sham79}
L. J. Sham and M. Nakayama, \prb {\bf 20}, 734 (1979).

\bibitem{koiller02}
B. Koiller, X. Hu, and S. Das Sarma, \prl {\bf 88}, 027903 (2002) 

\bibitem{schliemann01}
J. Schliemann, D. Loss, and A. H. MacDonald, \prb {\bf 63}, 085311 (2001).

\bibitem{hu02}
X. Hu and S. Das Sarma, \pra {\bf 66}, 012312 (2002).


\end{thebibliography}
\end{document}